\documentclass[runningheads]{llncs}
\usepackage{graphicx}

\usepackage[numbers]{natbib}

\usepackage{tabularx}
\usepackage{booktabs}

\usepackage{subfigure}

\usepackage{xcolor}

\newcommand\blfootnote[1]{%
  \begingroup
  \renewcommand\thefootnote{}\footnote{#1}%
  \addtocounter{footnote}{-1}%
  \endgroup
}

\usepackage{array}
\newcolumntype{x}[1]{>{\centering\arraybackslash\hspace{0pt}}p{#1}}

\definecolor{successColor}{HTML}{9bbb59}
\definecolor{failColor}{HTML}{bf504d}
\definecolor{blueLine}{HTML}{1111EE}

\begin{document}
\title{Tasks Decomposition Approaches in Crowdsourcing Software Development}
%
%\titlerunning{Abbreviated paper title}
% If the paper title is too long for the running head, you can set
% an abbreviated paper title here
%
\author{Abdullah Khanfor}
\authorrunning{A. Khanfor}
% First names are abbreviated in the running head.
% If there are more than two authors, 'et al.' is used.
%
\institute{College of Computer Science \& Information Systems, Najran University, Najran, Saudi Arabia
}
\maketitle              % typeset the header of the contribution
\begin{abstract}
A main characteristic of crowdsourcing software development (CSD) is the complexity of tasks and skills required by workers to achieve successful software crowdsourcing. The tasks proposed to the crowd in CSD are checked to ensure they are manageable and achievable. In general, individual tasks come from general goal-oriented projects. There are practices for breaking down software projects into manageable tasks, known as task decomposition. This study identified task decomposition techniques in software engineering, particularly in the context of CSD. Then, we defined the experienced developers who lead the requester in decomposing the project, preparing tasks, and reviewing submissions. This study explored and addressed decomposition approaches in CSD. Next, we selected projects in TopCoder to identify the task decomposition process in the CSD context. Finally, we concluded with future research directions for investigating decomposition approaches and their effects in the CSD context to ensure successful crowdsourced software projects.

\end{abstract}

\blfootnote{This paper is accepted for publication in the 25th International Conference on Human-Computer Interaction, Jul. 2023. \newline 
The Conference Proceedings will be published by \textcopyright~2023 Springer. Personal use of this material is permitted. Permission from Springer must be obtained for all other uses, in any current or future media, including reprinting/republishing this material for advertising or promotional purposes, creating new collective works, for resale or redistribution to servers or lists, or reuse of any copyrighted component of this work in other works.}
\section{Introduction}

The power and potential of crowdsourcing have been demonstrated in multiple fields using various statistical tools. Moreover, affordable, high-quality products and solutions should be created from diverse, experienced backgrounds to solve problems and fulfill the owner’s needs. ``Crowdsourcing is the act of taking a job traditionally performed by a designated agent (usually an employee) and outsourcing it to an undefined, generally large group of people in the form of an open call'' \cite{howe2006rise}. CSD is derived from crowdsourcing ideas through the proposal of software tasks such as developing, testing, and bug fixing in an open-call format \cite{shahin2015improving, mao2015developer}. In some circumstances, CSD is more appealing to developers and companies. CSD reduces costs and improves deliverable quality through reducing defect rates \cite{lakhani2010topcoder}.

\begin{figure}[h]
	\centering
	\includegraphics[width=0.9\textwidth]{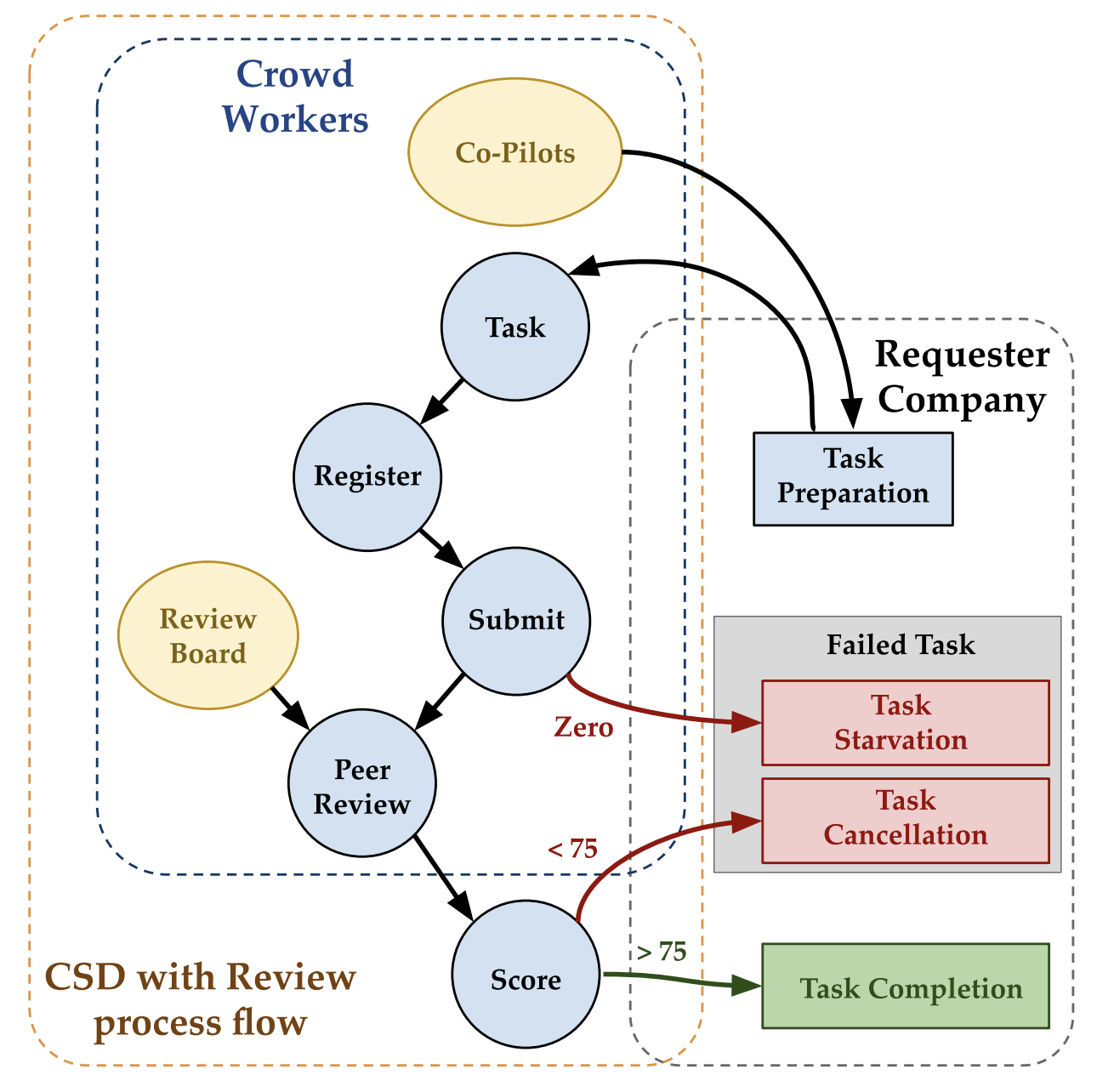}
	\caption{Copilots involvement in the general CSD workflow.}
	\label{FIG:1}
\end{figure}

On CSD platforms, the requester can ask for assistance in preparing and managing tasks from a pool of experienced workers. These experienced workers, known as copilots in TopCoder, improve the task description, reward, and communication with other workers to ensure the success of the task or project. Workers register for tasks to submit artifacts. Moreover, in TopCoder, submissions from the crowd are assigned scores out of 100 points. If there is no submission to the task, it is labeled as a failure, also known as task starvation \cite{khanfor2017failure}. If there are submissions with scores less than 75, the tasks are marked as a failure, and the task is canceled because of the low quality of the submissions. If the task submissions have a score of 75 or above, the top submissions are selected as winners, and a reward is given to the top submitters. Moreover, the copilot receives a separate award for guiding the requester and crowd to successfully completing the task. The CSD process using copilots is illustrated in Fig.~\ref{FIG:1}.

Various studies have examined the role of copilots in the CSD context. A significant role of copilots in CSD projects is task decomposition, which breaks down the project into a set of microtasks to ensure deliverable software quality. However, there are some conditions and trade-offs in choosing appropriate strategies for decomposing a project 
~\cite{szajnfarber2014exploring,lakhani2012open, saremi2019ant}. On the one hand, efficient task decomposition for a software project requires detailed specifications to capture the project’s requirements. On the other hand, it leaves room for innovation from the crowd \cite{lakhani2010topcoder}. Hence, The detailed decomposition can increase overhead communication, and coordination with workers \cite{latozashould}. Moreover, in addition to the accurate description of the task, decomposed tasks should have limited interdependences to ensure success \cite{tajedin2013determinants}. Therefore, methods for efficient task decomposition in the CSD context should be identified and examined. To this end, we identify approaches used in task decomposition to build a systematic approach that accelerates the CSD process and its adoption for software projects.

\section{Background}

This section examines software decomposition, task decomposition practices used in software engineering, and various decomposition terminologies, such as cohesion, coupling, and horizontal and vertical decomposition. Next, we highlight the main tasks of the copilot, such as task preparation and decomposition. The following is a background discussion of various software-task decomposition concepts and the role of the copilot in CSD.

\subsection{Software Decomposition}
The goal of software engineering decomposition is to reduce complexity  \cite{bruegge2010object}. The design of complex systems involves three critical structures \cite{parnas1984modular}. 1) The module structure assigns an individual programmer or team to closely related programs, where the module is a group of decomposed programs. 2) For the users of the structure, the components are part of the modules. 3) The process structure decomposes runtime activities, known as a process. There are no direct relationships between the modules and process, which is challenging to identify. During task decomposition, the following criteria must be considered when decomposing tasks. First, the duplication of tasks should be avoided. Second, the dependency of project tasks should be identified; for example, if \textit{Task 2} depends on the completion of \textit{Task 1}, this dependency must be maintained during task design. Finally, the generality concept in software development is when we can use other components in various situations without changes, which can be driven to task generality. Therefore, we can reuse other similar successful tasks for different projects in the CSD context.

\subsubsection{Cohesion \& Coupling}
The components of task decomposition must be cohesive and coupled. In task decomposition, a high cohesion between different tasks can indicate their similarity and vice versa. The increased coupling means a slight change in one subsystem can significantly impact other subsystems. The same concept can be applied to task decomposition. Thus, tasks with high coupling and dependency require adjustments to the other tasks.

\subsubsection{Horizontal \& Vertical Decomposition}
\begin{figure}[!t]
\centering
    \includegraphics[width=\textwidth]{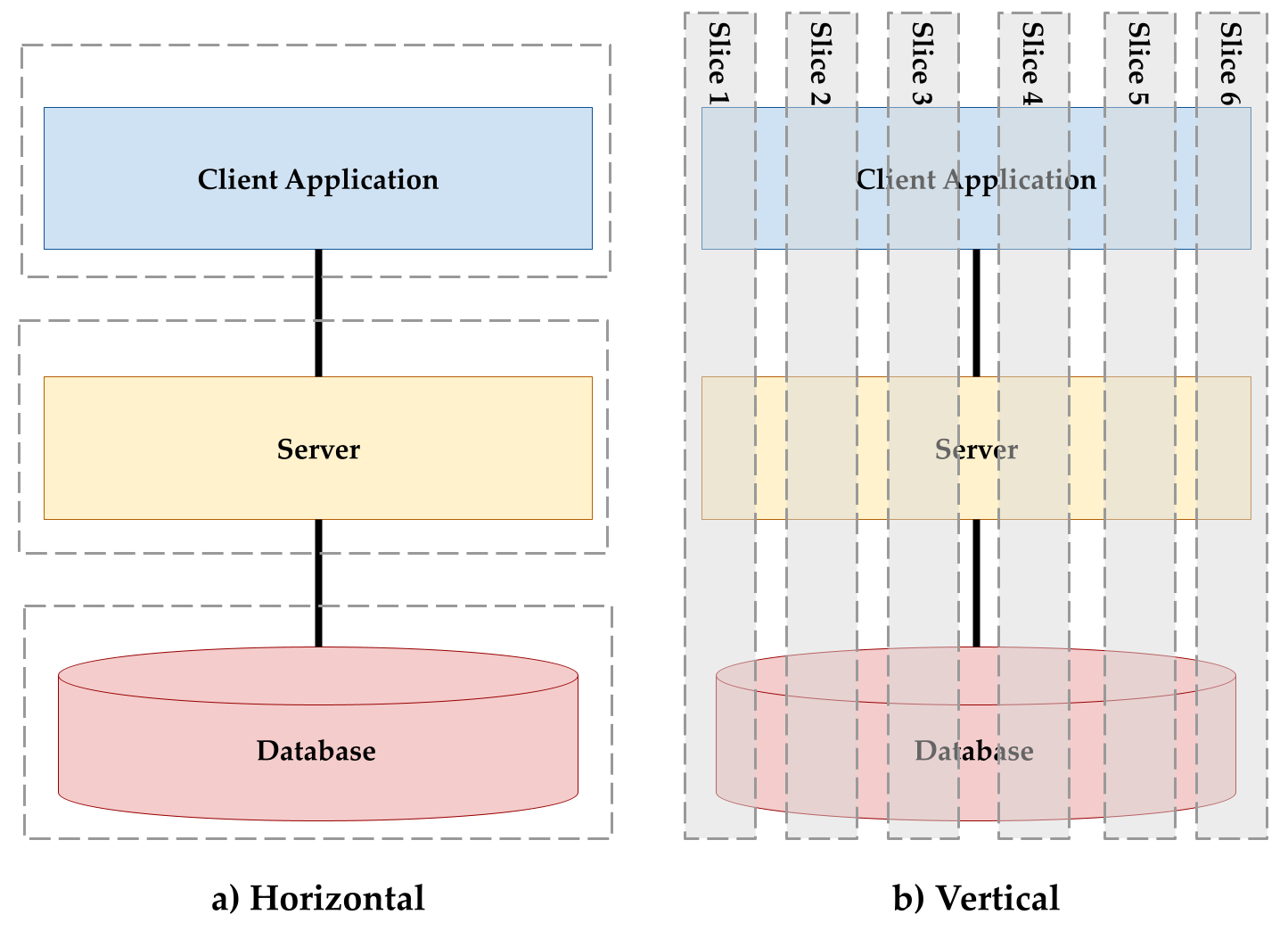}
    \caption{Client-server Architecture and Vertical Scales Demonstration \cite{adobe_split}.}
    \label{arc_slices}
\end{figure}

Decomposition in the Agile methodology can be divided into two main types: (1) horizontal and (2) vertical split \cite{adobe_split}. Horizontal split is the traditional approach in which each significant feature in the software is decomposed based on architectural layers. Therefore, the role of each layer may be assigned to a different individual or team. A significant benefit of this approach is the building of expertise among responsible specialists based on the assigned layer. Vertical decomposition involves decomposing a prominent feature that overlaps a different architectural layer. The breakdown satisfies the requirement that the development of each component, including construction and integration into the testing component, must take a few days. Green \cite{adobe_split} stated that the benefits of adopting the vertical decomposition approach could rely on a simple understanding, more accurate estimation, increasing the feedback loop, and noting the progress of a software project. Fig.~\ref{arc_slices} shows the vertical and horizontal slices of the client-server architecture.

\subsubsection{Decomposition Levels}
\begin{figure}[h]
	\centering
		\includegraphics[width=\linewidth]{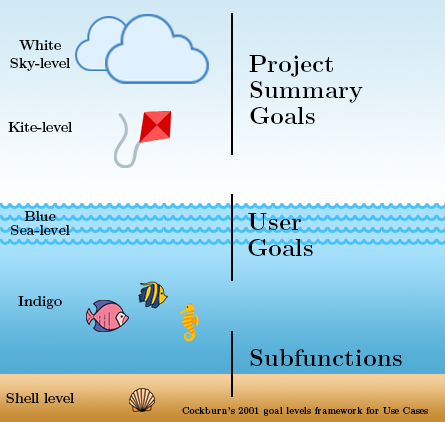}
	\caption{Illustration of different levels of requirements description.}
	\label{FIG:3}
\end{figure}

Each task should be labeled according to Cockburn’s altitude notation, Fig~\ref{FIG:3}, where tasks with too abstract goals are placed at the sky level, and software goals with no precise ending are placed at the kite level. At the kite level, CSD task decomposition is more intense. The greater detail of the functional and sea levels allows developers to complete the tasks in these levels in a few sittings. At the sub-functional level, each task may not individually indicate the goal of the software; however, multiple sub-functions can demonstrate the goal. Therefore, this level is the optimal level for tasks that can lead to higher parallelism and reduce the dependability of tasks. Finally, the deep-sea level or shell level should be more detailed and focus on the goal of the task, with less visibility of the project goals. In Fig~\ref{FIG:3}, illustration of different levels of requirements description. 

\subsection{Copilot roles in CSD}
Saengkhattiya et al. \cite{saengkhattiya2012quality} studied how quality is handled by crowdsourced companies. One such technique involves conducting interviews and questionnaires. They found that copilots generally act as “middlemen between the task requester and workers. The copiloting process includes discussions regarding various aspects of CSD, such as project planning, pricing, communicating, and writing task descriptions. Thus, the copilot faces considerable challenges that they must mitigate and resolve. The authors noticed several runs of testing and individual assessments performed by copilots that ensured high-quality worker submissions. Communication among different parties is a challenge in CSD \cite{stol2014two}. On the one hand, having one point of contact, the copilot, with whom the project owner can communicate is advantageous. On the other hand, this single line of communication inhibits direct communication between the workers and the project owner, especially in topics such as how to submit their solutions. Saengkhattiya et al. \cite{saengkhattiya2012quality} reported communication and coordination problems between customers and copilots; these communication problems lead to copilot changes initiated by the customer. The copilot can prevent the requester from hearing directly from the crowd community. In most cases, utilizing copilots increases the overall cost of the project.

A crucial role of the copilot is to prepare for task decomposition and identify dependencies among crowdsourced software projects. Farraj \cite{farrajcomponent} proposed concepts that can be employed on a software-driven automated copilot tool based on a predefined workflow; however, the empirical validation of the recommended tool was limited. The primary motivation for developing the tool was that copilots might unexpectedly leave a project or mismanage it, leading to its failure or delays.

TopCoder defines a copilot as a member who manages the TopCoder process for a customer to deliver the requested asset. The copilot controls the crowd's direction and is responsible for working with the customer at an appropriate level of detail to convert the business submission into technical requirements \cite{saengkhattiya2012quality}. Copilots earn money by successfully completing challenges in three main tracks: design, development, and data science. Copilots are part of the crowd that applies for this role and are certified after achieving a set of minimum eligibility requirements %\footnote{https://www.topcoder.com/thrive/articles/Copilot\%20&\%20Reviewer\%20Certification\%20Program}. 
The evaluation process can be divided into two parts: (1) pre-requisite requirements, including submitting and winning a specific number of challenges, in addition to having a certain number of posts in the TopCoder Forums, and (2) written exams, including writing challenge specifications, estimating budgets and timelines, and providing strategic feedback to customers. Applicants are evaluated based on the content of their answers and written English skills, which are necessary for clear communication with customers and the crowd. Furthermore, TopCoder continues to monitor copilot performance after the copilot is certified. While working with an actual customer, TopCoder members evaluate the copilot from many perspectives, including customer management, community management, technical skills, communication, and professionalism.

\begin{figure*}
	\centering
		\includegraphics[width=\textwidth]{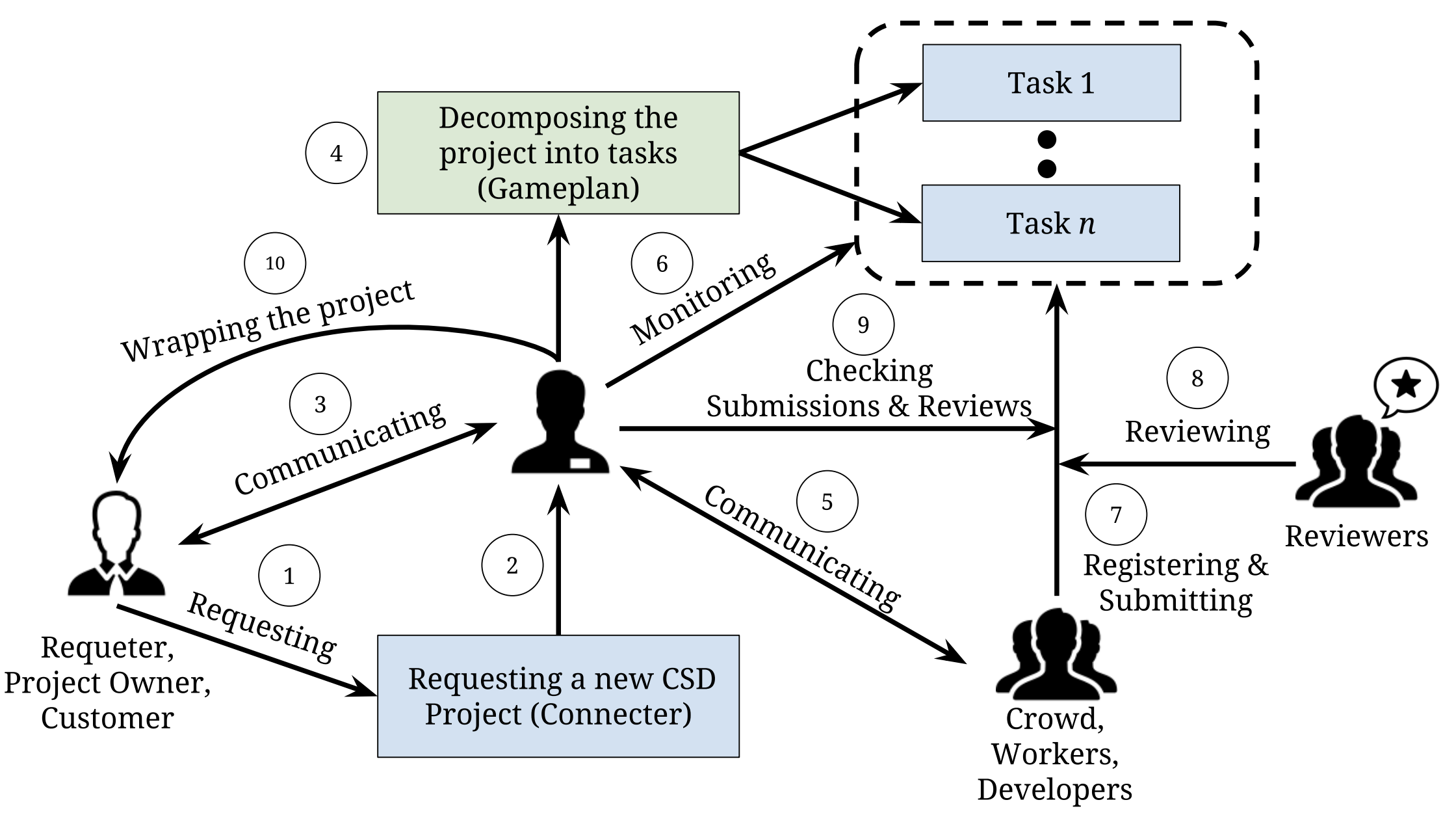}
	\caption{Copilot contribution in the CSD project workflow.}
	\label{FIG:2}
\end{figure*}

In Fig.~\ref{FIG:2}, we summarize the roles of the copilot during the project life cycle. In Step 1, the requester starts the project. During project initiation, the requester may select a copilot from a pool of copilots. Usually, copilots are matched by TopCoder (Platform) based on the utilization, availability, skills, and experience of the copilot for the new project. TopCoder creates a shortlist of possible copilots that satisfies previous criteria from which the project owner can select a copilot. Next, the selected copilot begins communicating with the project owner to clarify or discuss the project scope, plan, and requirements, as shown in Steps 2 and 3. The copilot takes the lead in creating tasks and the project timeline in Step 4, which decomposes different tasks for the crowd. After posting the task to the crowd in Step 5, the copilot communicates with workers to clarify the requirements of the created tasks. The copilot monitors the published tasks and ensures a high level of participation in Step 6. If required, the copilot may request the platform to announce the tasks to the crowd. The crowd performs the tasks, and after they submit their solutions, another set of crowd reviewers examines the deliverables of the tasks in Steps 7 and 8. Finally, in Step 9, the copilot reviews the submissions to ensure that the project is delivered successfully. In some cases, the workers may appeal to the review score, and the copilot analyzes the submissions and reviews to override the reviewer’s evaluation.

\subsection{Decomposition in Crowdsourced Tasks}
CrowdForge \cite{kittur2011crowdforge} proposed a framework to solve dependent and complex tasks using a map-reducing approach. Turkomatic \cite{kulkarni2011turkomatic} studied the design of the workflow of a task. It provides task decomposition that the requester can change. Another example is Cascade \cite{chilton2013cascade}, an automated object taxonomy that generates labels and selects the best one for categorization. Therefore, it creates highly parallelized mini-tasks that can be completed in 20 s. Despite these frameworks, their primary concern is with the MTurk task platform, which are short and low-reward tasks that do not require specific skills, unlike CSD Tasks. Therefore, their conclusions can be used but are implied in the CSD task scenario. Eickhoff and de Vries \cite{eickhoff2013increasing} found that a unique phrasing of the task description affects outcome quality. Thus, careful task decomposition that considers this finding will help provide a more robust framework that aids in task decomposition. Jiang and Matsubara \cite{jiang2014efficient} conducted a simulation to decompose a complex task into mini-tasks using two models. The authors analyzed vertical and horizontal task decomposition simulation results and found that, in general, vertical decomposition outperforms horizontal decomposition in obtaining better quality outcomes from the crowd.

\section{Methodology}

The study has two main objectives. First, identifying whether the decomposition approach of tasks in the historical CSD project was horizontal or vertical. To achieve this, we examined historical CSD tasks within projects using the technology required to accomplish the task. From there, if most of the tasks within a CSD project use both frontend and backend programming languages for each task, we conclude that the tasks were vertically decomposed; if a small percentage of tasks are denoted as both, and there is a more significant number of tasks that can be distinguished as frontend and backend, the tasks were horizontally decomposed. Second, we determine whether the projects selected for this study can be used to observe the effect of task decomposition on the success rate and overall success of CSD projects.

The historical dataset was collected from the TopCoder platform from January 2014 to January 2015. It contains 4907 tasks. Of the tasks, 84.3\% were marked as complete, indicating that the code submitted by the workers was rated, and the requester accepted it. The dataset contains the tasks and developer information. In each task, the associated metadata describe the task, including the task ID, title of the task, start dates for registration and submission, the end date of submissions, task type, and task description. Each task has several registered workers, and several submissions, and depending on whether it is completed, it is assigned a success or failure status. TopCoder defines a “completed” or a successful task as one with a submission with a score of at least 75 out of 100. If a submission has a score lower than 75 or there are zero submissions, the task is marked as ``canceled.'' \\

\section{Results}

We analyzed the types of tasks based on these technologies. We observed that there were 75 programming languages or frameworks in the proposed tasks, for example, PHP, iOS, and Node.JS. Of the identified technologies, 23 are used for front-end development, whereas 61 are used in back-end development. Note that 23 and 61 add up to more than 75 technologies, which is because some of the technologies can be used for both frontend and backend development. This helps us distinguish which type of decomposition is used for the tasks proposed in TopCoder. Fig.~\ref{pies} shows the percentages of frontend and backend tasks. From the figure below, we can conclude that Project $5104$ is horizontally decomposed. For Project $7424$, the frontend and backend tasks represent the highest percentage of tasks in the project. Therefore, this project was vertically decomposed.

\begin{figure}[!h]
\label{pies}
\centering
\includegraphics[width=\textwidth]{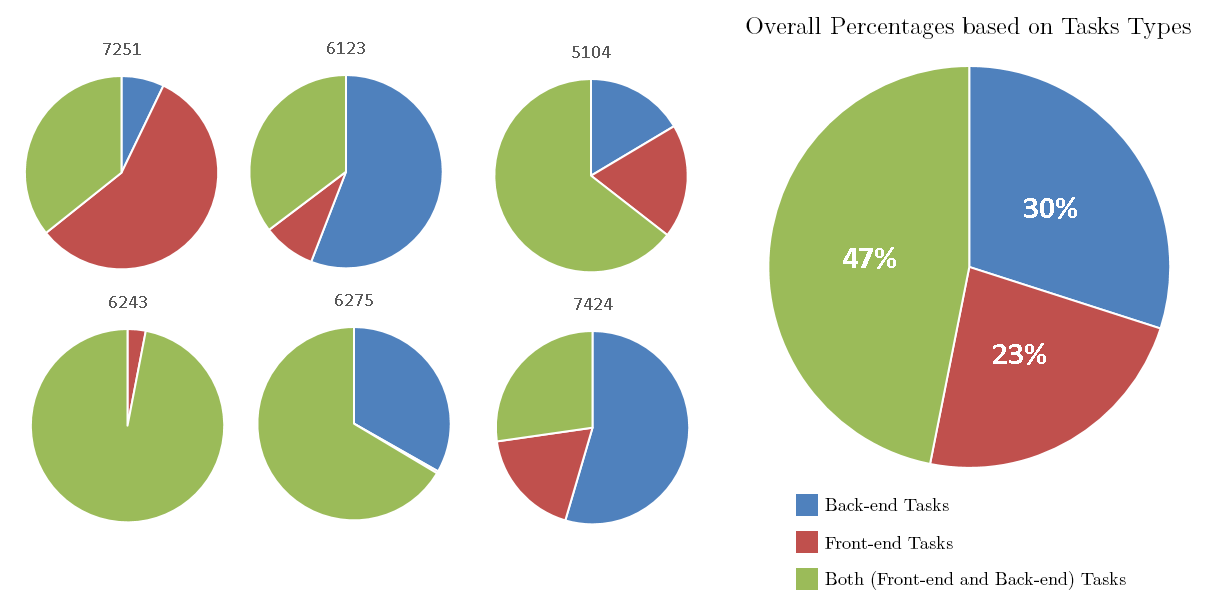}
\caption{Percentages of Task Type in 6 Projects and the Overall Dataset.}
\end{figure}

\begin{figure}[!h]
\label{fail_pre}
\centering
\includegraphics[width=\textwidth]{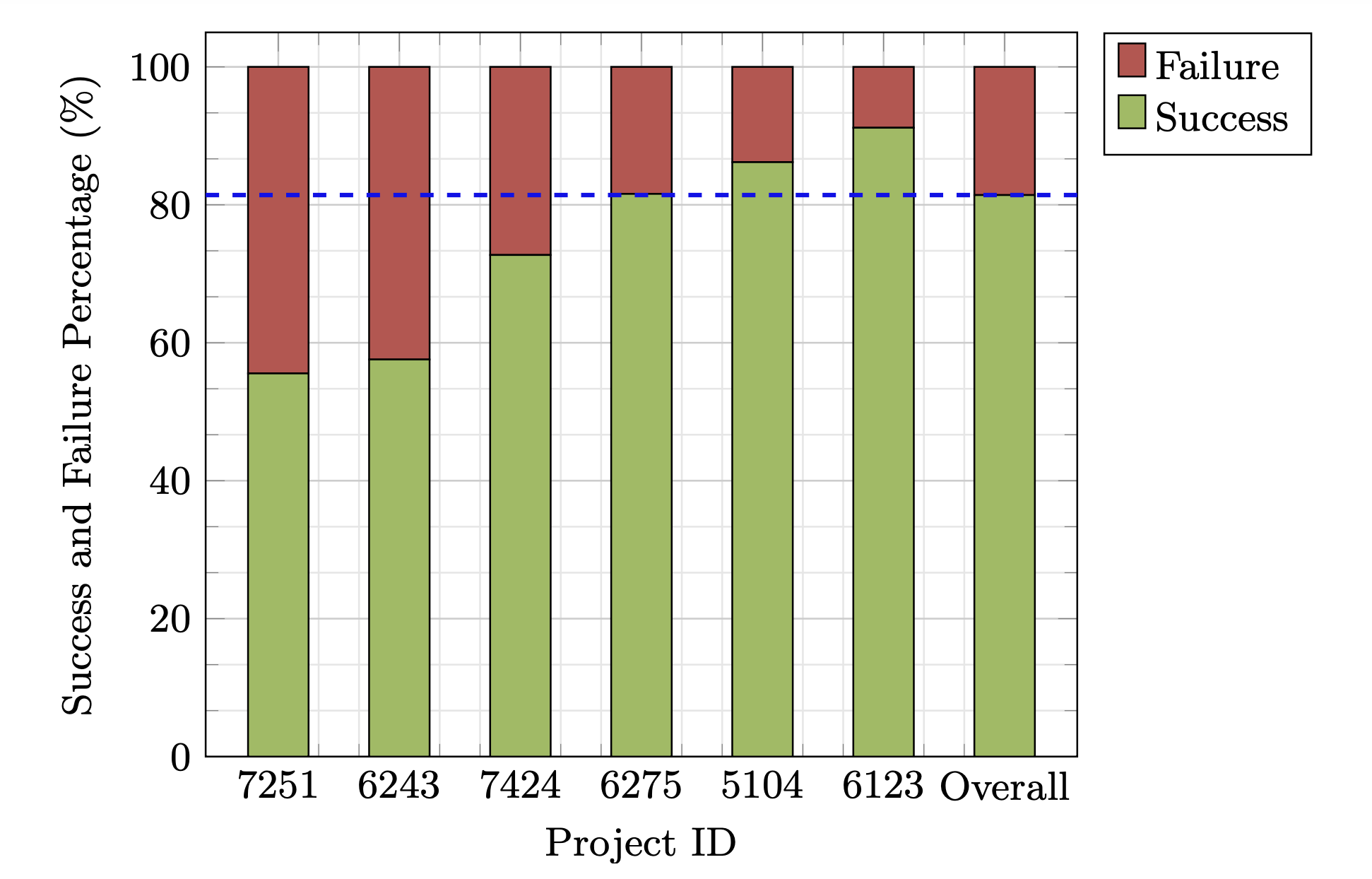}
\caption{Tasks failure rate for different projects and the overall dataset.}
\end{figure}

To examine the effect of the type of task decomposition on CSD, we used the failure rate of each project. The failure rate of the dataset was 15.7\%. Thus, if the project has a lower failure rate than the overall failure rate, it is assumed to be successful. For example, in Fig.~\ref{fail_pre}, we examine the selected projects, as stated, Project 5104 is horizontally decomposed and had a successful task rate. In contrast, project 7424, which is vertically decomposed, has a task success rate that is lower than the average overall success rate on the platform.

The award, complexity, duration, and number of workers in the task and their influence on the task’s success can help us identify which part of the CSD requires in-depth research \cite{khanfor2017failure}. The strong effect of task description on the task outcome acts as motivation for further research. Using the right task decomposition method can reduce the complexity of tasks by ensuring sufficient workload and price. In Fig.~\ref{fail_pre}, we observe that projects 7251, 6243, and 7424 have a higher failure rate than the overall failure rate of the tasks in the historical dataset, where the blue dashed line represents the acceptable failure rate of CSD projects. Meanwhile, projects 6275, 5104, and 6123 can be considered successful projects because they exceed the failure rate threshold (dashed line) of the referenced CSD projects.

%\begin{table}
%\centering
%\caption{Selected projects information about the decomposition approach and the failure rate.}\label{tbl_results}
%\begin{tabular}{cc c}
%\toprule
%Project ID & \hspace{0.25cm} Decomposition type \hspace{0.25cm} & Failure rate (\%) \\
%\midrule
%7251 & Vertical & 44.44 \\
%6243 & Horizontal & 42.42 \\
%7424 & Vertical & 27.27 \\
%6275 & Horizontal & 18.41 \\
%5104 & Horizontal & 13.81 \\
%6123 & Vertical & 8.823 \\
%\bottomrule
%\end{tabular}
%\end{table}

\section{Conclusion and Future Work}
This study investigated decomposition methods for software projects. First, we provide an overview of the role of copilots in CSD. A main task of a copilot is decomposing software projects into a set of tasks. Therefore, the decomposition techniques used in software engineering are reviewed. Next, we identified the vertical and horizontal decomposition methods. Then, we studied the decomposition of selected CSD projects. This study paves the way for investigating the importance of decomposing tasks to ensure the success of crowdsourced software projects. In future work, we will compare vertical and horizontal decompositions on a large scale and their impact on the success rate of CSD tasks.

\bibliographystyle{splncs04}
\bibliography{ref}

\end{document}